\documentclass[]{spie}  

\usepackage{amsmath,amsfonts,amssymb}
\usepackage{graphicx}
\usepackage[colorlinks=true, allcolors=blue]{hyperref}
\usepackage{booktabs}
\usepackage[utf8]{inputenc}
\usepackage{graphicx}
\usepackage{subcaption}
\usepackage{array}
\usepackage{multirow}
\usepackage{xcolor}
\usepackage{siunitx}
\usepackage{hyperref}

\definecolor{spiegold}{RGB}{200,168,0}
\definecolor{rowgray}{RGB}{245,245,245}
\definecolor{rowwarn}{RGB}{255,235,200}
\definecolor{negred}{RGB}{180,0,0}
\definecolor{posgreen}{RGB}{0,100,0}

\newcolumntype{C}[1]{>{\centering\arraybackslash}p{#1}}
\newcolumntype{L}[1]{>{\raggedright\arraybackslash}p{#1}}

\newcommand{\mum}{\,\si{\micro\meter}}

\usepackage{tikz,xcolor,hyperref}
\definecolor{lime}{HTML}{A6CE39}
\DeclareRobustCommand{\orcidicon}{%
    \begin{tikzpicture}
    \draw[lime, fill=lime] (0,0) 
    circle [radius=0.16] 
    node[white] {{\fontfamily{qag}\selectfont \tiny ID}};
    \draw[white, fill=white] (-0.0625,0.095) 
    circle [radius=0.007];
    \end{tikzpicture}
    \hspace{-2mm}
}
\newcommand{\orcid}[1]{\href{https://orcid.org/#1}{\orcidicon}}

\title{Determining stress-based bending mode limits for the\\
       Vera C. Rubin Observatory M1M3 active mirror system}

\author[a]{Malhar Sonaniskar\orcid{0000-0002-4573-3027}}
\author[a]{Douglas Neill}
\author[a]{Ellie Hileman}
\author[b]{Petr Kubánek\orcid{0000-0002-1877-1386}}

\affil[a]{NSF-DOE Vera C. Rubin Observatory/NSF-NOIRLab, Tucson, AZ 85719, USA}
\affil[b]{NSF-DOE Vera C. Rubin Observatory/NSF-NOIRLab, La Serena, Chile}

\authorinfo{Further author information: (Send correspondence to Malhar Sonaniskar)\\Malhar Sonaniskar: E-mail: malhar.sonaniskar@noirlab.edu, Telephone: 1 520 318 8402}

\begin{document}
\maketitle

\begin{abstract}
The Vera C. Rubin Observatory Simonyi Survey Telescope's primary–tertiary mirror (M1M3) is an actively supported, 8.4-m cast borosilicate optic controlled by 156 pneumatic actuators. This work presents a rapid stress-estimation methodology based on the root-sum-square (RSS) combination of Finite Element Analysis to derive pre-computed unit bending mode stresses. Since the stress is proportional to strain, and strain is proportional to displacements, we theorized that since the bending mode displacements can be combined RSS, that the peak stresses would also combine by RSS. We validate the RSS-based major principal stress predictions against NASTRAN simulations for representative bending mode combinations, demonstrating agreement within a few percent for peak Principal major stress across the mirror glass substrate. Unit displacement and corresponding unit stress fields for the first 20 natural bending modes of the M1M3 system are generated using NASTRAN. Representative multi-mode corrections including combinations that include astigmatism, coma, and spherical modes of higher order are then analyzed to compare the resulting peak principal stresses with RSS-based predictions. The method enables near instantaneous evaluation of stress margins for active optics corrections, safety-limit checking, and actuator-force optimization during telescope operations. This paper outlines the formulation, implementation workflow, validation results, and practical use cases for integrating RSS-based stress prediction into the Vera C. Rubin Observatory's M1M3 active optics system.
\end{abstract}

\keywords{Finite Element Analysis, Mirror bending modes, RSS stress prediction, Stress}

\section{INTRODUCTION}
\label{sec:introduction}

Large astronomical telescopes rely on actively controlled primary mirrors to maintain optical performance under varying gravitational, thermal, and operational loads. The Vera C. Rubin Observatory features a monolithic primary-tertiary mirror (M1M3) fabricated from borosilicate glass with an overall diameter of 8.4~m.\cite{sebag2016,neill2016cell} The mirror incorporates an active support system composed of 156 pneumatic force actuators (FA) and six hardpoints.\cite{neill2016cell,daruich2018} The FA provides necessary support forces, including corrective force in the axial and lateral direction as necessary to offload all gravity loads, effectively "floating" the mirror in space. A hexapod of axially stiff hardpoints kinematically define the M1M3 position in tip, tilt, and piston. The mirror's active optical figure corrections are coordinated through the Active Optics System (AOS), which applies force corrections derived from wavefront measurements.\cite{neill2014aos,thomas2016} During telescope operations, actuator forces are continuously adjusted to correct optical aberrations arising from gravity deformation, thermal gradients, and mechanical disturbances.\cite{daruich2018,xin2020} While actuator commands are primarily derived from optical surface measurements, it is equally necessary to ensure that force configurations do not induce excessive mechanical stresses within the mirror glass substrate. The M1M3 is a 17 metric-ton monolithic borosilicate structure that is irreplaceable within operational timescales, therefore, glass safety is a critical constraint on every actuator command.\cite{quint2024,daruich2024}

Finite Element Analysis (FEA) models provide accurate stress predictions but require substantial computational time, making them incompatible with real-time control decisions. This limitation led to the development of rapid predictive methods capable of estimating mirror stresses without performing full FEA simulations for each candidate correction vector. This paper introduces a stress prediction technique based on the Root-Sum-Square (RSS) superposition of pre-computed unit normalized bending mode stresses. The method enables nearly instantaneous peak stress estimation using pre-computed finite element solutions and is validated against NASTRAN FEA results for correction trials.

\section{BACKGROUND AND THEORY}
\label{sec:theory}

The structural response of the M1M3 mirror to actuator loading is governed by the equations of linear elasticity. Within this framework, stress is proportional to strain, and strain is proportional to displacement, a chain of proportionality fundamental to the RSS stress prediction methodology. Because the mirror glass substrate remains well within its elastic regime during all active optics corrections, the superposition principal applies: the total stress field resulting from any combination of actuator force sets is the sum of the stress fields produced by each force set acting independently.\cite{timoshenko1970}

The M1M3 interfaces with the Rubin AOS to operate primarily from a look-up table (LUT) supplemented by closed-loop corrections derived from curvature wavefront sensors.\cite{neill2014aos,thomas2016} Closed-loop corrections are evaluated and applied at minute timescale, after each science exposure is read out and processed. The optical surface figure corrections are decomposed into a linear combination of bending modes, and the required actuator forces are determined by the corresponding modal correction coefficients.\cite{xin2020,blomquist2023} For each bending mode $i$ with correction coefficient $c_i$, the peak stress contribution scales as $c_i \cdot \sigma_i^{\mathrm{unit}}$, where $\sigma_i^{\mathrm{unit}}$ is the pre-computed peak major Principal stress for mode $i$, normalized to 1\,\si{\micro\meter} RMS optical surface displacement. Glass principally fails by brittle facture. Its strength is limited by the major tensile stress, given by the peak major principal stress. Because bending mode displacements combine by RSS, a consequence of their orthogonality in displacement space, the peak stresses are hypothesized to combine by RSS as well.

Optical testing of the fully assembled M1M3 mirror and support system at the University of Arizona Richard~F.\ Caris Mirror Lab confirmed that the differences between measured bending modes and FEA-predicted modes are less than a few percent.\cite{xin2020} This validates the fidelity of the FEA model used in the present work for deriving unit bending mode stress fields.

The RSS method is expected to be most accurate when modal stress fields are spatially distributed across different mirror regions. The method may underpredict when two or more modes produce large stresses at the same element location, since those stresses add linearly in FEA but in quadrature under RSS. Rotationally symmetric modes, the Focus/Spherical (Mode~3) and 2nd Spherical (Mode~12) shape, share the same radial peak stress locations by definition and represent the highest risk for underprediction when combined with large simultaneous coefficients.

\section{FINITE ELEMENT MODEL}
\label{sec:fem}

The M1M3 FEA model was developed and refined in support of optical testing performed at the University of Arizona Richard~F.\ Caris Mirror Lab.\cite{xin2020,sebag2016} The M1M3 monolithic substrate is a cast borosilicate hexagonal honeycomb sandwich fabricated at the Mirror Lab.\cite{martin2016,sebag2016} It has a continuous planar backplate and facesheet connected by ribs in a hexagonal pattern, with 426 steel pucks bonded to the backface and connected through invar loadspreaders to the 156 pneumatic force actuators.\cite{neill2016cell}

The FEA model captures the accurate mass and stiffness properties of the borosilicate glass mirror substrate. The glass mirror body is discretized using plate (shell) elements, with rod elements representing the load spreaders and actuator stroke lengths. A total of 156 nodes representing the actuator locations were placed according to the as-built drawings of the M1M3 mirror cell support system.\cite{neill2016cell} The mirror cell structure has been analyzed in detail to confirm that deck deflections remain within limits that do not overstress the mirror under gravity and seismic loading.\cite{neill2016cell}

The mirror is constrained using a three-point kinematic constraint method. All three constraint nodes are restrained in the axial (Tz) direction, two in the lateral (Ty) direction, and one in (Tx). Reaction forces at these constraint nodes were verified to be negligible after each static analysis, confirming that the hardpoints do not influence actuator forces, surface deformation, or glass stress under any of the load cases examined.

Modal analysis of the complete FEA model was performed to extract the natural bending modes of the glass mirror on its active support. The first six modes correspond to rigid body motions (tip, tilt, piston, Z rotation, and X-Y lateral translations) and are excluded from further analysis. Modes 7 through 26 in the NASTRAN modal ordering, relabeled as Bending Modes~1 through~20 in this work, represent the first 20 elastic deformation modes and form the correction basis for this study. These 20 modes collectively represent the correction content most relevant to active optics operations, higher modes are rarely excited at significant amplitude under normal observing conditions.\cite{xin2020,neill2014aos} These higher modes are not utilized in the current closed-loop system.

\begin{figure}[htbp]
    \centering
    \includegraphics[width=0.65\linewidth]{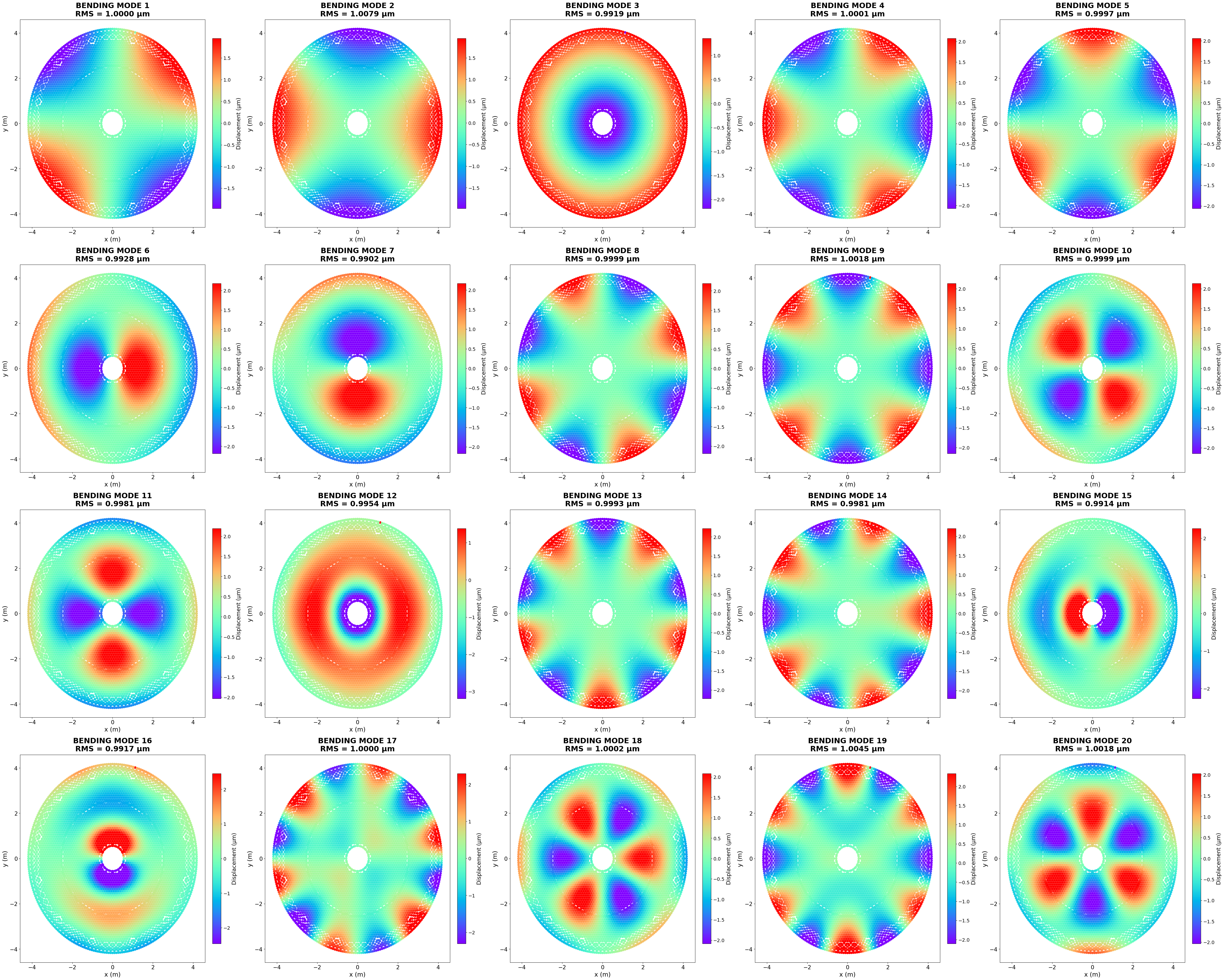}
    \caption{M1M3 bending modes normalized to 1\,\si{\micro\meter}~RMS. Tip, tilt, and piston have been removed by least squares plane fitting. Orthogonal conjugate pairs are visible throughout. Mode~3 (Focus/Spherical) and Mode~12 (2nd Spherical) are the two rotationally symmetric modes and represent the highest risk for RSS underprediction when combined with large simultaneous coefficients.}\label{fig:modes}
\end{figure}

For each of the 20 elastic bending modes, optical surface deformation is extracted at 5,256 nodes on the top face of the M1M3 mirror and normalized to an RMS amplitude of 1\,\si{\micro\meter} along the optical axis. Tip, tilt, and piston contributions are removed from all surface error plots using a least-squares plane fitting procedure implemented in Python. The resulting normalized actuator force matrix has dimensions $20 \times 156$. Figure~\ref{fig:modes} shows the optical surface deformation plots for all 20 positive bending modes at 1\,\si{\micro\meter}~RMS normalization. 

\section{RSS METHODOLOGY AND IMPLEMENTATION}
\label{sec:method}

\subsection{Unit bending mode stress computation}

For each of the 20 elastic bending modes, the normalized actuator forces corresponding to a 1\,\si{\micro\meter}~RMS optical surface correction are applied to the FEA model in a separate static analysis. The Major Principal (tensile) stress, and the Minor Principal (compressive) stress at every element centroid in the mirror substrate is extracted, and the peak value across all elements is stored as $\sigma_i^{(+)}$ for the positive sense of mode $i$, and $\sigma_i^{(-)}$ for the negative sense.  Since glass fails by brittle fracture, its strength is limited by the major principal stress.  However, each bending mode has a inverse negative value. Since the system is linear, for the negative value the  Major and Minor Principal stresses invert, and minor principal stress becomes the limiting major principal stress.  This procedure yields 40 pre-computed unit peak stress values stored as a static LUT. Centroidal stress values are used throughout, as they are directly available from NASTRAN output without additional post-processing.

\vspace{\baselineskip}
Table~\ref{tab:unit_stress} presents the full set of unit Major Principal stresses for all 20 bending modes. Stress magnitude generally increases with mode number, reflecting the higher spatial-frequency deformations of higher order modes. 

\begin{table}[htbp]
\centering
\small
\begin{tabular}{L{2.5cm} C{3.5cm} C{3.5cm}}
\toprule
\textbf{Bending Mode} &
\textbf{Major\ Prin.\ Stress (+) [MPa]} &
\textbf{Minor\ Prin.\ Stress ($-$) [MPa]} \\
\midrule
Mode 1  & $1.62\times10^{-2}$ & $-1.60\times10^{-2}$ \\
Mode 2  & $1.61\times10^{-2}$ & $-2.03\times10^{-2}$ \\
Mode 3  & $4.72\times10^{-2}$ & $-2.20\times10^{-2}$ \\
Mode 4  & $4.84\times10^{-2}$ & $-4.85\times10^{-2}$ \\
Mode 5  & $4.35\times10^{-2}$ & $-4.33\times10^{-2}$ \\
Mode 6  & $7.61\times10^{-2}$ & $-7.58\times10^{-2}$ \\
Mode 7  & $6.05\times10^{-2}$ & $-6.04\times10^{-2}$ \\
Mode 8  & $9.33\times10^{-2}$ & $-9.44\times10^{-2}$ \\
Mode 9  & $8.46\times10^{-2}$ & $-7.80\times10^{-2}$ \\
Mode 10 & $8.36\times10^{-2}$ & $-8.32\times10^{-2}$ \\
Mode 11 & $7.53\times10^{-2}$ & $-8.43\times10^{-2}$ \\
Mode 12 & $2.07\times10^{-1}$ & $-1.03\times10^{-1}$ \\
Mode 13 & $2.50\times10^{-1}$ & $-2.51\times10^{-1}$ \\
Mode 14 & $2.27\times10^{-1}$ & $-2.27\times10^{-1}$ \\
Mode 15 & $3.22\times10^{-1}$ & $-3.22\times10^{-1}$ \\
Mode 16 & $3.99\times10^{-1}$ & $-3.99\times10^{-1}$ \\
Mode 17 & $5.44\times10^{-1}$ & $-5.42\times10^{-1}$ \\
Mode 18 & $1.68\times10^{-1}$ & $-1.66\times10^{-1}$ \\
Mode 19 & $5.24\times10^{-1}$ & $-2.39\times10^{-1}$ \\
Mode 20 & $1.44\times10^{-1}$ & $-1.42\times10^{-1}$ \\
\bottomrule
\end{tabular}
\caption{Major and Minor Principal stress M1M3 bending modes 1-20 - normalized to 1µm RMS optical surface error}
\label{tab:unit_stress}
\end{table}

\subsection{RSS prediction formula}

Given a set of bending mode correction coefficients $\{c_i\}$ for modes $i = 1, \ldots, 20$, the predicted peak glass tensile stress is: 
\begin{equation}
\sigma_{\mathrm{RSS}} = \sqrt{\sum_{i=1}^{20} \bigl(c_i \cdot\sigma_i^{\mathrm{unit}}\bigr)^2}
\label{eq:rss}
\end{equation} where for each mode, $\sigma_i^{(+)}$ is used when $c_i > 0$ and $\sigma_i^{(-)}$ when $c_i < 0$. This formula requires only 40 multiplications, 40 squarings, a summation, and a square root, rapidly calculated on any modern control computer. 

\subsection{Comparison with direct summation} For reference, a direct summation estimator is defined as:
\begin{equation}
\sigma_{\mathrm{sum}} = \sum_{i=1}^{20} \bigl|c_i \cdot \sigma_i^{\mathrm{unit}}\bigr|
\label{eq:sum}
\end{equation}
This assumes worst-case constructive interference that all mode stress peaks occur at the same element simultaneously. As shown in Sec.~\ref{sec:validation}, this assumption leads to severe and unrealistic overprediction (101--302\% above FEA), making summation unsuitable for operational stress monitoring.

\subsection{Operational integration}

The RSS stress estimate is computed immediately after the AOS wavefront estimation pipeline delivers a new correction vector and before actuator commands are transmitted. The M1M3 support system control operates an inner loop controlling individual actuator forces and an outer loop computing the figure correction commands from wavefront measurements.\cite{daruich2018,quint2024} The RSS check integrates naturally at the outer loop boundary: if $\sigma_{\mathrm{RSS}}$ exceeds the operational glass stress threshold, the correction is flagged for review before being dispatched to the pneumatic force actuators.

The method also enables stress-constrained actuator force optimization at negligible computational cost, allowing optimization algorithms to incorporate stress constraints without iterative FEA calls. The 40-element pre-computed unit stress LUT requires no updates unless the FEA is revised to reflect as-built changes to the mirror or support system. For reference, the M1M3 support system has demonstrated robust quasi-static and dynamic performance during commissioning with a steel surrogate mirror, and full glass mirror testing is now underway.\cite{quint2024,daruich2024}

\section{VALIDATION}
\label{sec:validation}

\subsection{Initial trial set: all modes combined}

\begin{figure}[htbp]
    \centering
    \includegraphics[width=0.8\linewidth]{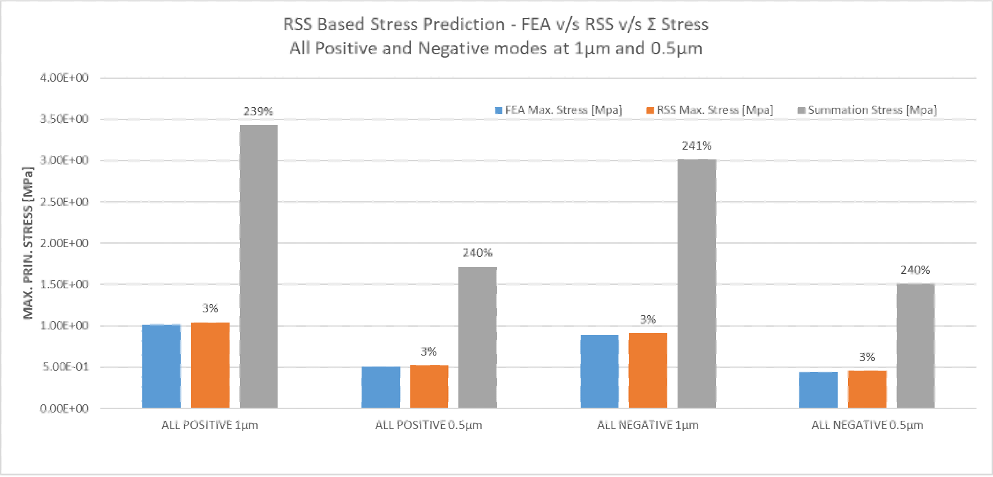}\caption{RSS-based peak stress prediction vs.\ FEA for initial trial of all bending modes positive/negative applied simultaneously. Percentage labels indicate difference relative to FEA.}\label{fig:Initial_Trial_Results}
\end{figure}

As a first validation, all 20 bending modes were activated simultaneously, both all-positive and all-negative, at uniform correction coefficients of 1\,\si{\micro\meter} and 0.5\,\si{\micro\meter}~RMS/mode. These extreme, non-realistic corrections distribute large energy equally across all modes and serve to establish baseline RSS performance under maximum loading. For all four cases, RSS predicts FEA stress to within $+3\%$, as shown in Figure~\ref{fig:Initial_Trial_Results}. In  contrast, direct summation overestimates FEA stress by 239--241\% (a factor of approximately $3.4\times$), confirming the spatial independence of M1M3 bending mode stress fields across the full 20-mode basis.

\subsection{Realistic bending mode combination trials}

\begin{figure}[htbp]
    \centering
    \includegraphics[width=1.0\linewidth]{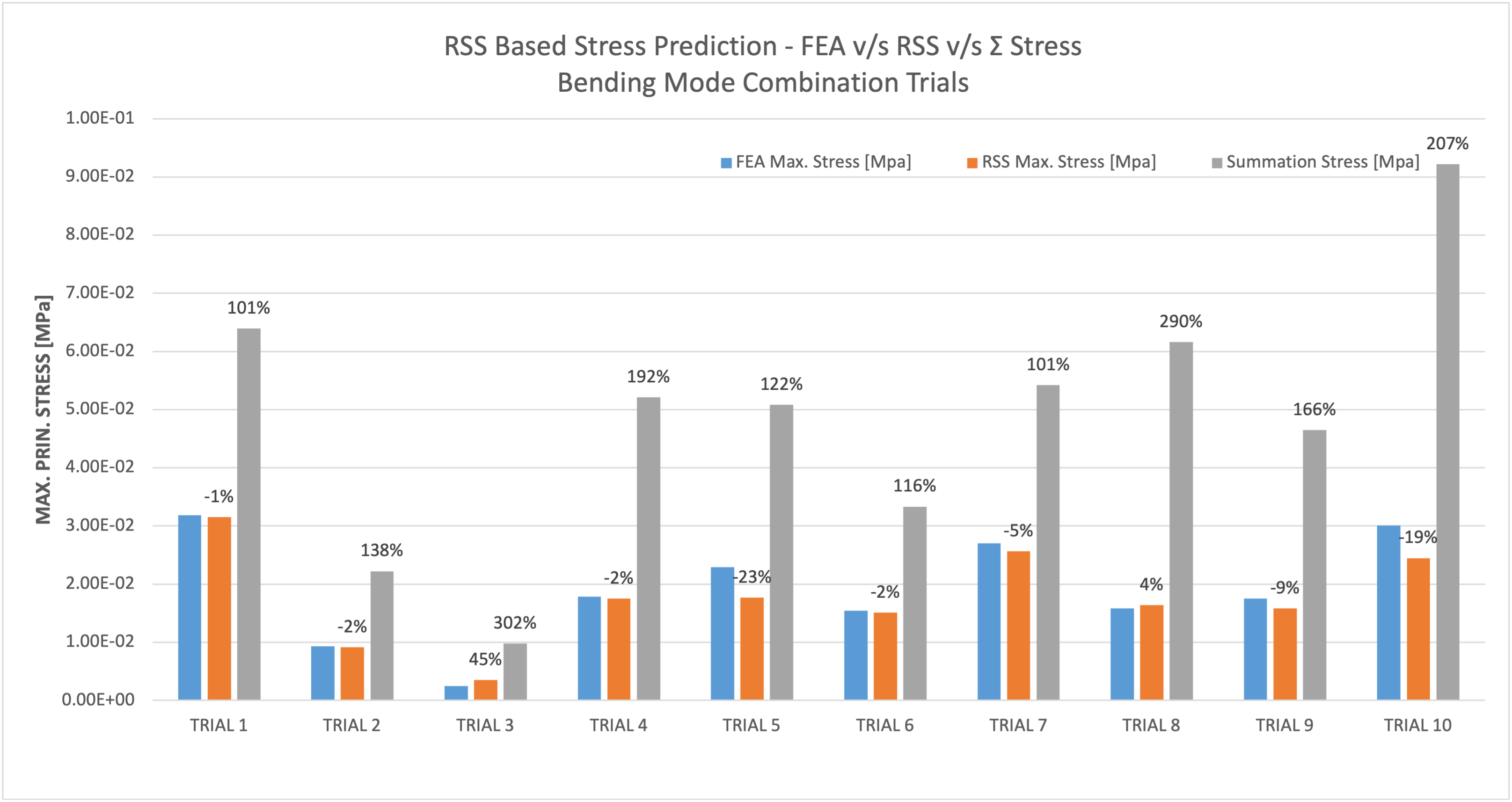}\caption{RSS-based peak stress prediction vs.\ direct FEA for the ten realistic bending mode combination trials. Percentage labels indicate difference relative to FEA.}\label{fig:Realistic_Trial_Results}
\end{figure}

Ten realistic multi-mode correction trials (Trials~1--10) were constructed spanning sparse single-dominant corrections through dense multi-mode scenarios with coefficients following a $1/\mathrm{mode}^2$ amplitude falloff. For each trial, the complete coefficient set was applied in a single NASTRAN static analysis, and the peak centroidal Major Principal stress was extracted as FEA ground truth.  Table~\ref{tab:validation} presents the complete numerical comparison. Figure~\ref{fig:Realistic_Trial_Results} shows the bar chart comparison. 

\begin{table}[htbp]
\centering
\small
\begin{center}
\label{tab:trial_blend}
\begin{tabular}{|c|l|c|c|c|c|}
\hline
\rule[-1ex]{0pt}{3.5ex}
\textbf{Trial} &
\textbf{Category} &
$N_{\mathrm{pos}}$ &
$N_{\mathrm{neg}}$ &
\textbf{Max(+) [\si{\micro\meter}]} &
\textbf{Max($-$) [\si{\micro\meter}]} \\
\hline
\rule[-1ex]{0pt}{3.5ex}
T01 & Single dominant -- Focus & 1  & 2  & 0.0055 & 1.9486 \\
\hline
\rule[-1ex]{0pt}{3.5ex}
T02 & Single dominant -- Astigmatism & 1  & 3  & 0.0025 & 0.4777 \\
\hline
\rule[-1ex]{0pt}{3.5ex}
T03 & Sparse low-order (3--5 modes) & 2  & 3  & 0.1002 & 0.0501 \\
\hline
\rule[-1ex]{0pt}{3.5ex}
T04 & Sparse low-order (3--5 modes) & 2  & 3  & 0.3394 & 0.8162 \\
\hline
\rule[-1ex]{0pt}{3.5ex}
T05 & Sparse mixed (6--8 modes) & 2  & 5  & 0.0241 & 0.9354 \\
\hline
\rule[-1ex]{0pt}{3.5ex}
T06 & Sparse mixed (6--8 modes) & 4  & 6  & 0.2115 & 0.9353 \\
\hline
\rule[-1ex]{0pt}{3.5ex}
T07 & Dense -- 12 modes & 5  & 7  & 1.4027 & 0.1570 \\
\hline
\rule[-1ex]{0pt}{3.5ex}
T08 & Dense -- 14 modes & 10 & 7  & 0.3873 & 0.7585 \\
\hline
\rule[-1ex]{0pt}{3.5ex}
T09 & All-negative sparse & 0  & 6  & 0.0000 & 0.7347 \\
\hline
\rule[-1ex]{0pt}{3.5ex}
T10 & Full-spectrum dense & 11 & 11 & 0.1054 & 1.2087 \\
\hline
\end{tabular}
\caption{Bending mode combination trial set summary. $N_{\mathrm{pos}}$ and
$N_{\mathrm{neg}}$ denote the number of active positive and negative bending mode
coefficients, respectively. Maximum coefficient magnitudes are in \si{\micro\meter}~RMS.}
\end{center}
\end{table}

\begin{table}[htbp]
\centering
\small
\begin{tabular}{L{2cm} C{2cm} C{2cm} C{1.5cm} C{2cm} C{2.2cm} C{1.5cm}}
\toprule
\multirow{2}{*}{\textbf{Case}} &
\multicolumn{3}{c}{\textbf{FEA vs.\ RSS}} &
\multicolumn{3}{c}{\textbf{FEA vs.\ Summation}} \\
\cmidrule(lr){2-4}\cmidrule(lr){5-7}
& FEA [MPa] & RSS [MPa] & $\Delta\%$ & FEA [MPa] & Sum.\ [MPa] & $\Delta\%$ \\
\midrule
Trial 1  & 3.19E-02 & 3.15E-02 & $-1\%$  & 3.19E-02 & 6.40E-02 & \textcolor{negred}{101\%} \\
Trial 2  & 9.34E-03 & 9.13E-03 & $-2\%$  & 9.34E-03 & 2.22E-02 & \textcolor{negred}{138\%} \\
Trial 3  & 2.43E-03 & 3.51E-03 & \textcolor{posgreen}{$+45\%$} & 2.43E-03 & 9.75E-03 & \textcolor{negred}{302\%} \\
Trial 4  & 1.79E-02 & 1.75E-02 & $-2\%$  & 1.79E-02 & 5.21E-02 & \textcolor{negred}{192\%} \\
Trial 5  & 2.29E-02 & 1.77E-02 & \textcolor{negred}{$-23\%$} & 2.29E-02 & 5.09E-02 & \textcolor{negred}{122\%} \\
Trial 6  & 1.54E-02 & 1.51E-02 & $-2\%$  & 1.54E-02 & 3.33E-02 & \textcolor{negred}{116\%} \\
Trial 7  & 2.70E-02 & 2.56E-02 & $-5\%$  & 2.70E-02 & 5.42E-02 & \textcolor{negred}{101\%} \\
Trial 8  & 1.58E-02 & 1.64E-02 & $+4\%$  & 1.58E-02 & 6.16E-02 & \textcolor{negred}{290\%} \\
Trial 9  & 1.75E-02 & 1.58E-02 & $-9\%$  & 1.75E-02 & 4.65E-02 & \textcolor{negred}{166\%} \\
Trial 10 & 3.01E-02 & 2.45E-02 & \textcolor{negred}{$-19\%$} & 3.01E-02 & 9.22E-02 & \textcolor{negred}{207\%} \\
\midrule
All (+) 1\mum  & 1.01E+00 & 1.04E+00 & $+3\%$ & 1.01E+00 & 3.43E+00 & \textcolor{negred}{239\%} \\
All (+) 0.5\mum& 5.04E-01 & 5.21E-01 & $+3\%$ & 5.04E-01 & 1.72E+00 & \textcolor{negred}{240\%} \\
All ($-$) 1\mum  & 8.86E-01 & 9.11E-01 & $+3\%$ & 8.86E-01 & 3.02E+00 & \textcolor{negred}{241\%} \\
All ($-$) 0.5\mum& 4.44E-01 & 4.56E-01 & $+3\%$ & 4.44E-01 & 1.51E+00 & \textcolor{negred}{240\%} \\
\bottomrule
\end{tabular}
\caption{Validation results: FEA vs.\ RSS and FEA vs.\ direct summation for all ten realistic trials and four all-mode initial trial cases. $\Delta\%$ is computed as $(\text{prediction} - \text{FEA})/\text{FEA} \times 100\%$.}
\label{tab:validation}
\end{table}

\subsection{Analysis of results}

RSS agreement with the FEA based major principal stress is within $\pm5\%$ for 7 of 10 realistic trials. Trial~1 achieves near-perfect agreement ($-1\%$) because it is dominated by a single bending mode. When one mode carries the great majority of the RSS energy budget, the prediction reduces to the unit stress scaled by the correction coefficient, which is exact by construction.

Trial~3 shows a $+45\%$ overprediction because the correction is very sparse with a small number of active modes at low amplitude. The activated modes produce stress fields that partially cancel spatially, FEA captures this cancellation, while RSS adds all contributions positively in quadrature, producing a conservative overestimate. Trials~5 and~10 show $-23\%$ and $-19\%$ underpredictions, respectively. These cases contain mid-order modes with substantial coefficients whose spatial stress patterns partially overlap. When multiple modes produce large stresses at the same element location, those stresses add linearly in FEA but only in quadrature under RSS.

In contrast, direct summation overestimates FEA stress by 101--302\% across all ten realistic trials. This severe overprediction renders summation operationally useless for stress monitoring: it would require impossibly conservative force limits to stay within a summation-based stress budget. The RSS method provides a far more physically realistic estimate.

\subsection{Application of a constant multiplicative factor}

The RSS method shows promising results for implementation in the M1M3 AOS as a rapid prediction tool for glass stresses corresponding to any given bending mode combination and its associated corrective forces. While the RSS method occasionally underpredicts the FEA result, particularly when spatially correlated mode pairs dominate the correction, this limitation can be effectively addressed through the application of a constant multiplicative safety factor. Applying a factor of $1.25\times$ to the RSS-predicted stress threshold before comparison with the allowable glass stress limit ensures a conservative bound across the complete range of trials tested. With this factor applied, the RSS-corrected predictions yield results that are substantially better than the summation method in every case (summation overestimates by 101--302\%) while remaining conservative with respect to the FEA ground truth. This combination, RSS with a modest safety factor, provides an operationally viable approach to real-time stress monitoring that is both physically meaningful and computationally instantaneous.

\section{USE CASES AND OPERATIONAL INTEGRATION}
\label{sec:usecases}

The RSS stress prediction method provides three distinct operational capabilities for M1M3 support system and AOS control management:

\textbf{Real-time safety checking by support system control.} The RSS estimate is computed immediately upon receipt of a new correction vector from the AOS wavefront pipeline, before actuator commands are dispatched to the 156 pneumatic force actuators.\cite{daruich2018} If $\sigma_{\mathrm{RSS}}$ (adjusted by the calibration factor) exceeds the operational glass stress limit, the correction is flagged and the control system or operator can modify or reject it before any force is applied to the mirror.

\textbf{Bending mode stress optimization by the closed-loop feedback control.} The RSS formula can be incorporated as a stress constraint in the actuator force optimization loop. Because RSS evaluation is computationally negligible compared to FEA, stress-constrained optimization can proceed iteratively without any FEA calls, enabling the AOS to find force solutions that simultaneously satisfy optical figure requirements and glass stress limits.

\textbf{Stress minimization of the LUT.} The pre-computed 40-element unit stress LUT enables rapid stress surveys across the full accessible correction coefficient space, supporting construction of stress-versus-performance trade curves and establishment of mode-by-mode amplitude limits for LUT design across all telescope pointing configurations.\cite{quint2024}

\section{CONCLUSIONS AND FUTURE WORK}
\label{sec:conclusions}

This paper presents and validates a rapid RSS-based stress prediction methodology for the LSST M1M3 active mirror system.
The key findings are:

\begin{enumerate}
\item RSS combination of pre-computed unit bending mode stresses predicts peak Major Principal stress to within $\pm5\%$ for physically representative multi-mode corrections, and within $-23\%$ to $+45\%$ across the full trial set. A constant multiplicative safety factor of $1.25\times$ applied to the RSS threshold provides a conservative operational bound across all tested cases.

\item Applied with this calibration factor, the RSS method yields results that are significantly more accurate and physically meaningful than direct modal summation, which overestimates FEA stress by 101--302\% and is operationally unworkable as a safety monitor.

\item For all-mode corrections at uniform amplitude, RSS matches FEA within $3\%$, validating the spatial orthogonality assumption across the full 20-mode basis.

\item Single-dominant corrections achieve near-perfect RSS accuracy by construction, as the prediction reduces to the unit stress scaled by the correction coefficient.

\item The method requires only a 40-element pre-computed LUT and evaluates in microseconds, enabling near realtime stress monitoring and safety-limit checking at full telescope control cadence.
\end{enumerate}

The next scope of this research is to utilize actual bending mode combination data from the Rubin Observatory. Extending this work, a parallel effort will apply the same RSS stress prediction framework to the Rubin Observatory M2 secondary mirror, establishing a successful setup for rapid stress prediction across the full telescope active optics system. This will also serve to demonstrate the generalization of the RSS approach over different large optical geometries and glass compositions, broadening the potential applicability to other observatory systems.

\acknowledgments 
We acknowledge the support and extend our sincere thanks to the optics and mechanical team who were involved with the optical testing of the Vera C. Rubin's M1M3 mirror design and manufacturing. Their meticulous work on developing CAD and FEA models enabled us to continuously test and optimize our hypothesis. The Vera C. Rubin Observatory is funded by the National Science Foundation (NSF) and the U.S. Department of Energy (DOE), their generous support has been instrumental in advancing this groundbreaking project.

Malhar Sonaniskar has used Gemini AI to prepare large set of potential bending mode combinations based on the publicly released information of M1M3's geometry and support system. These were further refined to combination trials with higher probability of occurrence during scientific operation of the Vera C. Rubin Observatory. Additional AI help available in Overleaf was used for text correction.

\bibliography{report} 
\bibliographystyle{spiebib} 

\end{document}